\documentclass[a4paper, amsfonts, amssymb, amsmath, reprint, showkeys, nofootinbib, twoside, longbibliography]{revtex4-1}
\usepackage[english]{babel}
\usepackage[utf8]{inputenc}
\usepackage[colorinlistoftodos, color=green!40, prependcaption]{todonotes}
\usepackage{amsthm}
\usepackage{mathtools}
\usepackage{physics}
\usepackage{xcolor}
\usepackage{graphicx}
\usepackage[left=23mm,right=13mm,top=35mm,columnsep=15pt]{geometry} 
\usepackage{adjustbox}
\usepackage{placeins}
\usepackage[T1]{fontenc}
\usepackage{lipsum}
\usepackage{csquotes}
\usepackage{soul}
\usepackage[pdftex, pdftitle={Article}, pdfauthor={Author}]{hyperref} 
\usepackage{enumitem}

\begin{document}
\title{A Future-Input Dependent model for Greenberger-Horne-Zeilinger correlations}

\author{Izhar Neder}
    \email[Correspondence email address: ]{izhar.neder@gmail.com}
    \affiliation{Soreq Nuclear Research Center, Yavne 81800, Israel}
\author{Nathan Argaman}
    \email[Correspondence email address: ]{argaman@mailaps.org}
    \affiliation{Department of Physics, Ben-Gurion University of the Negev, Be'er Sheva, Israel and Department of Physics, Nuclear Research Center -- Negev, P.O. Box 9001, Be'er Sheva 84190, Israel}
\date{\today} 

\begin{abstract}
It is widely appreciated, due to Bell's theorem, that quantum phenomena are inconsistent with local-realist models.  In this context, locality refers to local causality, and there is thus an open possibility for reproducing the quantum predictions with models which internally violate the causal arrow of time, while otherwise adhering to the relevant locality condition.  So far, this possibility has been demonstrated only at a toy-model level, and only for systems involving one or two spins (or photons).  The present work extends one of these models to quantum correlations between three or more spins which are entangled in the Greenberger-Horne-Zeilinger state.
\end{abstract}
\maketitle

\section{Introduction}

\begin{figure}
    \centering
    \includegraphics[scale=0.95]{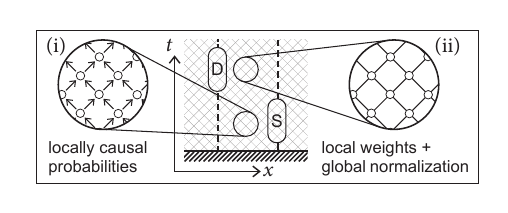}
\caption{A sketch of a spacetime region described by a mathematical model (center panel), with marks for a source (S), a detector (D) and initial conditions (hatched).  The blowups indicate two possibilities for the relations between model internal variables: (i) a locally causal description, with arrows indicating probabilistic dependencies (Bayesian network); (ii) a time-symmetric local description (Gibbs field), as per Eq.~\eqref{eq:P(I,O,U)} below.} 
\label{fig:blowups}
\end{figure}

It is well known that quantum mechanics (QM) violates ``local realism'' \cite{EPR1935}.  Indeed, Bell's theorem rules out local-hidden-variable models of quantum entanglement, under a broad set of assumptions which may be taken to represent the scientific method \cite{bell2004}.  These assumptions include the identification of variables which may be externally controlled in the lab as mathematical free variables (barring ``superdeterminism'' \cite{thooft2016}); the acceptance of irreversibly-recorded results of experiments as constraints on reality  (barring many-worlds approaches \cite{waegell2020}); and a causal arrow-of-time assumption, barring so-called retrocausal approaches \cite{price1997}.  However, the last assumption is too restrictive, as it excludes established scientific approaches such as the stationary action principle or Feynman's path integrals \cite{wharton2015a}.  Indeed, in order to evaluate and minimize the classical action, both initial and final conditions must be supplied as inputs, and thus the procedure is future-input dependent (FID), a term introduced in \cite{wharton2020}.  Such all-at-once (block universe) approaches are more time-reversal symmetric than a standard description, which explicitly begins with initial conditions and proceeds sequentially in time, see Fig.~\ref{fig:blowups}.

The acceptance of the arrow of time as a basic background assumption allowed Bell to demonstrate that any mathematical model reproducing the predictions of QM must employ non-local variables, in a certain precisely-defined sense \cite{bell1990}.  This, of course includes the standard quantum description, where the state of one particle of an entangled pair, $A$, changes as a result of performing a measurement on its distant partner, $B$ (the state of $A$ after the measurement depends on the result obtained for $B$).  On the other hand, QM meets many other locality conditions, and in particular signal locality (the probabilities of any outcome for measurements on $A$ do not depend on any external influence acted on B, e.g., what measurement was made on $B$).  As signaling involves only inputs and outputs, any model reproducing the input-output relations predicted by QM will also exhibit signal locality.

This tension between locality for signals and non-locality for the internal variables is difficult to swallow, but one is forced to accept it due to Bell's theorem and the subsequent experimental work \cite{hill2022}.  Similarly, when one considers reproducing the predictions of QM with FID models, signal causality will be maintained despite the fact that the internal variables of the model depend on inputs from the future, simply because agreement with QM is required.  On the other hand, the main difficulty with nonlocality, modeling connections between distant regions of space-time, e.g., through a state in an abstract Hilbert-space, need not feature in such models.  In particular, the influences between different variables may be completely local in an appropriate sense (see below).


We thus limit attention to ``local FID models'' which reproduce at least some of the results of QM, and fit the following general description.  The spacetime region covered by the model is discretized into a set of zones, such that the notion of neighboring zones is well-defined, and each of the model variables or parameters is associated with one such location.  The model describes the physical system considered in terms of local input parameters, $I_n\in I$, specifying the setup of the system and the measuring apparatus (these are the same input variables required by QM to make predictions for the system). Similarly, the model uses local output parameters $O_n \in O$, describing the measurement outcomes, and (optionally) local internal parameters $U_n \in U$.  The model specifies, for each set of values of the inputs, a distribution function over the internal and the output variables $P_I(O,U)$ such that the marginal distribution $P_I(O)=\sum_{U}{P_I(O,U)}$ agrees with the results of QM for the same physical system, at least in an appropriate limit. A continuous generalization may also be considered for the zones and the variables, where the summation sign is replaced by integration for continuous variables (also note that the use of probability distributions does not preclude a discussion of deterministic models, which merely correspond to the limit of very sharp distributions).

It is further required that the model defines local weight functions $w_l$, each of which depends on subsets of the model parameters $I_l \subset I$, $O_l \subset O$ and $U_l \subset U$ corresponding to a limited vicinity in spacetime.  The probability distribution $P_I(O,U)$ of the model is specified by the normalized product of these weights $w_l$,
\begin{equation}
P_I(O,U) = \frac{1}{Z} \prod_l w_l(I_l,O_l,U_l),   \label{eq:P(I,O,U)}
\end{equation}
where $Z=\sum_{O,U} \left( \prod_l w_l \right)$ (this normalization sum is over all possible combinations of values of all the parameters in $O$ and $U$, just as in statistical mechanics; indeed, such a random-variable model is known in the literature as a Gibbs field).  
That the $w_l$ are local, implies that the subsets $I_l, O_l, U_l$ are limited to one zone, or to a few adjacent zones.    Loosely speaking, they are limited to the "$l$th neighbourhood." 

Finding a general local FID model which agrees with a wide range of predictions of QM remains a Grand Challenge. Importantly, there do exist such descriptions which reproduce the two-particle correlations of QM employed in proofs of Bell's Theorem. We focus on one of these, Schulman's L\'evy-flight model, which is described in detail below (see \textcite{wharton2020} for a review). 
In this brief article, another step toward a local FID model as a possible reformulation of QM is reported: We 
show that Schulman's model with a minor modification can quantitatively describe Greenberger-Horne-Zeilinger (GHZ) three-particle correlations \cite{GHZ,GHSZ1990}.  It can also describe its N-particle generalization.

We first briefly present the GHZ setup (Sec.~II) and Schulman's L\'evy-flight toy model (Sec.~III), and show intuitively how the latter can be naturally generalized to describe the former (Sec.~IV).  We then construct a quantum circuit that generates the GHZ correlations for $N$ particles, and show explicitly how locality is maintained 
in the model, by means of the local quantum gates of the circuits (Sec.~V).  We end with a concluding discussion (Sec. VI).

\section{Three-spin entanglement}
\label{sec:GHZ}

One may discuss the GHZ setup by considering the following state of three spin-$\frac{1}{2}$ particles:
\begin{equation}
    |\Psi \rangle_{\rm GHZ} = \frac{1}{\sqrt{2}} \left( | \uparrow\uparrow\uparrow \rangle 
+ | \downarrow\downarrow\downarrow \rangle \right)
\end{equation}
(the up and down arrows denote spin in the $\pm z$ direction).  It is assumed that the particles are prepared in this state by a source at some early time, and are later transported to well-separated locations, where spin measurements may be performed.  Consider measurements of the $\theta_j$ spin component of each particle, where the angle $\theta_j$ is in the $x-y$ plane and is measured from the $x$ axis (here $j=1,2,3$ distinguishes between the particles).  The operators associated with these measurements are 
$\hat \sigma_j = \left( \begin{matrix} 0 & \exp(-i\theta_j) \\  \exp(i\theta_j)  & 0 \\ \end{matrix}\right)$ (a factor of $\hbar/2$ is dropped), with eigenvalues $A_j = \pm 1$ and corresponding eigenvectors $\frac{1}{\sqrt{2}} \left( \begin{matrix} 1 \\ A_j\exp(i\theta_j) \\ \end{matrix}\right)$.  Using $N=3$ for the number of spins, the probability amplitude for obtaining the set of results $\{ A_j \}_{j=1..N}$ is $\frac{1}{\sqrt{2^{N+1}}} \left[ 1+ \left(\prod_{j=1}^N A_j \right) \exp \left( i \sum_{j=1}^N \theta_j \right) \right]$, and the probability is
\begin{equation} \label{P_QM}
P_{\rm GHZ} \left( \{A_j\} \right) =
\frac{1}{2^N} \left[ 1+ \left(\prod_{j=1}^N A_j \right)\cos \left( \sum_{j=1}^N \theta_j \right) \right] .
\end{equation}
As can be seen by just changing $N$ above, this formula holds as well for more or less than three particles, a result to be used below. With this distribution, one finds that the expectation value of the product of all the spin measurements is $\left\langle \prod_j \hat \sigma_j \right\rangle = \left\langle \prod_j A_j \right\rangle = \cos \left( \sum_j \theta_j \right)$ where the products and sums run over $j=1,2,\dots N$. 

With this state, demonstrating the incompatibility of QM with local causality is easier than with the usual proof of Bell's theorem, as no inequalities are needed \cite{GHSZ1990,mermin1990}.  It suffices to consider measurements of the $x$ and $y$ components, so that $\theta_j$ is equal to either $0$ or $\pi/2$ for each $j$, and to limit attention to either a measurement of all the $x$ components ($\theta_j=0$) or measurements in which only one of the particles has the $x$ component of its spin measured, and the other two particles have the $y$ component of their spin measured ($\sum_j \theta_j = \pi$).  In these cases, while the individual results are uncertain, with probabilities of $50\%$ for either $A_j = 1$ or $A_j = -1$ for each $j$, the product of the measurement results, $\prod_j A_j$, is predicted by QM with certainty: it is $+1$ if only $x$ components are measured, and $-1$ in the cases involving $y$ measurements.


This situation, in which the measurement outcomes are maximally correlated (or anticorrelated) is just as considered by Einstein, Podolsky and Rosen in \cite{EPR1935}. 
For example: the outcome of the measurements of the $y$ component of the second spin, $A_2$, is fully determined if one knows the other measurement outcomes of the $x$ component of the other  two spins,  $A_1$  and $A_3$, as their product $A_1A_2A_3$ must be $-1$.  As the measurements are to be made at space-like separations, a locally causal description
would necessarily have
to provide a predetermined value for $s_{2y}$, 
regardless of whether the $x$ component of spins 1 and 3 are actually measured or not.  In fact, it would have to provide predetermined values for 
each one of the possible spin measurements --- $s_{1x}$, $s_{2x}$, $s_{3x}$, $s_{1y}$, $s_{2y}$ and $s_{3y}$.
These predicted $s = \pm 1$ values of the different spins may be described as being set according to any joint rule by the source, but these $s$ values must be fixed once the particles have separated sufficiently from each other, and before the measurement settings $\theta_j$ can have any effect on them.


However, \textit{no set} of such preexisting values for the results of the spin measurements can be compatible with the predictions of QM, as that would require $s_{1x} s_{2x} s_{3x} = 1$ and $s_{1x} s_{2y} s_{3y} = 
 s_{1y} s_{2x} s_{3y} = s_{1y} s_{2y} s_{3x} = -1$. This is an impossible combination, as the product of these four predictions of QM is $\prod_{j}{s_{jx}^2s_{jy}^2}=-1$, but this product of squares is $+1$, non-negative. 
 This contradiction provides a proof of Bell's Theorem --- no locally causal model or description can reproduce the predictions of QM --- without recourse to expectation values and inequalities.
 Below we show that a simple local FID model can produce these predictions, as well as the full probability distribution for the general set of values $\left\{ 0 \leq \theta_j<2\pi \right\}$, Eq.~(\ref{P_QM}).

\section{Schulman's L\'evy-flight model }

We now turn to Schulman's model (see section~9.1 of \textcite{schulman1997} for the single-particle case, and \textcite{wharton2014,wharton2016} for the two-particle scenario relevant to Bell's Theorem; see also Ref.~\cite{wharton2020}, section IV.B).  In this FID model, the spin of the different particles is described by a time-dependent unit vector, $\vec q_j(t)$.  In the simple examples to be considered here, these vectors lie in the $x$-$y$ plane, and so can be identified with angles $\varphi_j(t)$ from the $x$ axis.  The model prescribes probabilities for the overall histories (or configurations) consisting of all the values of the $\vec q_j$s at all times and all the measurement results $A_j$ involved in each setup, subject to boundary conditions specified by the preparation procedures and the measurements involved.

In the limit relevant to achieving agreement with QM, the rules of the model are as follows.  The vector $\vec q_j(t)$ is constant in time for each particle, except for the possibility of a kink at some intermediate time $t_k$.  As few kinks as possible occur, as required for consistency with the boundary conditions of the model: if there are solutions with no kinks, the kinks become irrelevant; if there are none, and there are solutions with a single kink occurring at some time $t_k$, then solutions with two or more kinks are irrelevant, etc. 
 This is a limiting case of the type of random walk called a L\'evy flight, and can be achieved with local rules by taking the weight for no change in the orientation $\varphi_j(t)$ at any given time very much larger than the weights for a kink (see Sec.~V A below for details). 
 
Denoting the values of $\vec q_j(t)$ at times just after and just before $t_k$ by $\vec q_{j\pm}$, the probabilities are determined by the weight of such a kink event, $w_k$, which is taken to be proportional to  $\cos^2(\Delta \varphi/2 )$ (or $1+\cos{\Delta\varphi}$) where $\Delta \varphi=\varphi_{j+}-\varphi_{j-}$ is the change in the spin direction ($\cos{\Delta\varphi}=\vec q_{j-}\cdot\vec q_{j+}$).  A constant factor in the weights is immaterial, 
as Eq.~(\ref{eq:P(I,O,U)}) includes an appropriate normalization procedure.  The rules for measurements specify that $\vec q_j$ at the time of measurement points either in the direction of $\theta_j$ or in the opposite direction, i.e., $\varphi_j=\theta_j$ or $\varphi_j=\theta_j+\pi$ at the time of the measurement, with $A_j=1$ in the first case and $A_j=-1$ in the second 
It remains to specify the rules for the preparations.

For the single-spin situation, the initial state is a spin in some given orientation $\varphi_0$ (the subscript $j$ is omitted).  The model correctly reproduces the probabilities of QM for a subsequent spin measurement, as the piecewise-constancy of $\vec q(t)$ implies that $\vec q_{-}$ is determined by $\varphi_0$ and $\vec q_{+}$ is determined by $\theta$ and the outcome $A = \pm 1$, and the  rules specified above for a kink are precisely those necessary to reproduce the predictions of QM: one finds $\Delta \varphi_{1} = \theta-\varphi_0$ for the outcome $A=1$, and $\Delta \varphi_{-1} = \theta-\varphi_0 + \pi$ for the outcome $A=-1$, with the probability  $\cos^2(\Delta \varphi_{A}/2)$ for each case.  Note that for cases in which QM makes predictions with certainty, such as a spin initially pointing in the $+x$ direction and measured in the $-x$ direction, there is no need for a kink. 
Multiple sequential measurements on a single spin are also clearly covered by the model \cite{leggett1985}, with repeated measurements in the same orientation requiring no additional kinks (of course, an extra index is then required on $A_{i}$ and $\Delta \varphi_{i}$).

For two spins in the symmetric Bell state $\frac{1}{\sqrt{2}} \left( | \uparrow\downarrow \rangle 
- | \downarrow\uparrow \rangle \right)$ the preparation specifies that the spin vectors of the two particles are opposite at the source, so $\vec q_{2-} = -\vec q_{1-}$ (using $j = 1,2$), with no bias on the actual direction, see \textcite{wharton2014}.  In this case a single kink anywhere along the world lines of both particles suffices to satisfy any boundary conditions resulting from the future measurements, and the model reproduces the predictions of QM for two-spin Bell correlations: $P_{\theta_1,\theta_2}(A_1,A_2) =
\frac{1}{4} \left[ 1 - A_1A_2 \cos ( \theta_1 - \theta_2) \right]$ .  
For two particles with spins measured in the same direction, there are two possibilities which do not require a kink: $A_1=-A_2=1$ and $A_1=-A_2=-1$.  The model associates equal probabilities for these two possible outcomes, $50\%$ each.

The different possibilities for the timing of the kink and for which spin is affected by the kink all lead to the same results in this type of model.  Different variants of the model treat this issue differently.  \textcite{wharton2014} ascribes equal probabilities to all possible timings along the paths, whereas the possibility that the kink always happens within the region defined by the source is promoted in section 10.2 of \textcite{schulman1997}, as this resolves the issue of conservation of angular momentum.  For the present purposes, all the different variants are legitimate toy models for quantum correlations.
See also \textcite{maudlin1994,wood2015} for counterarguments to these models, and \textcite{berkovitz2008,almada2016} for corresponding rebuttals. 

\section{A local FID model for GHZ correlations}\label{sec:ourFid}

This section generalizes the FID model described above for one and two spins to the three-spin GHZ situation.  More details, including a generalization to $N$ spins, will be given in the next. 

An appropriate initial condition for the GHZ state specifies that all the $\vec{q_j}$s are in the $x$-$y$ plane, with 
correlated orientations such that 
\begin{equation}    
\varphi_{1-}+\varphi_{2-}+\varphi_{3-} = 0
\end{equation}
(equality for angles is modulo $2\pi$). There is no other restriction on the directions due to the initial setup. Again, at most a single kink anywhere along the world lines of all three particles suffices to reproduce the probabilities of QM.  At the time of the measurement of each spin, their outcomes $A_j\in\{+1,-1\}$ imply that the angles of the $\vec q_{j+}$ are
\begin{equation}
    \varphi_{j+}(t) = \theta_j + \pi (1-A_j)/2 \, ,
    \label{eq:varphi_j+}
\end{equation}
and these values hold fixed from the time of the kink, $t_k$, up to the time of the measurement.  Note that there is a kink only for one value of $j$; for the others $\varphi_{j+}=\varphi_{j-}$.

Similarly to the FID model for one and two spins, the probabilities of QM are reproduced by the weight
\begin{equation}
w_k(\Delta\varphi) = \cos^2\left(\frac12\Delta\varphi\right)
\end{equation}
associated with the kink at which one of the angles changes by $\Delta\varphi$. Note that due to the initial condition, we have after the kink $\sum_j \varphi_{j+} = \Delta\varphi$. The probability to find a particular experimental outcome  $\{A_1,A_2,A_3\}$ is therefore proportional to 
\begin{align} 
P_{\rm FID}\left( \left\{ A_j\right\}\right) & \propto
\cos^2\left(\frac{1}{2} \sum_j \varphi_{j+}\right) \nonumber \\ & = \cos^2\left( \frac{1}{2} \sum_{j=1}^3 \left[ \theta_j + \frac{\pi}{2} (1-A_j) \right] \right) \, .
\label{P_GZH_lv}
\end{align} 
As $A_j\in\{1,-1\}$, and there are eight possible combinations of the three outcomes, Eq.~\ref{P_GZH_lv} is identical to the result of QM, given by Eq.~\ref{P_QM}.  Note that the expression above is also valid when no kink is necessary, i.e., when $\sum_j \theta_j$ is a whole-integer multiple of $\pi$ --- the possibility of no kink at all is seamlessly reproduced by the expressions for a kink with $\Delta\varphi \to 0$.
The essential message is that due to FID, there is no insurmountable difficulty in reproducing the predictions for entangled systems, in a model which is local in the appropriate sense.





The examples used in the proof of Bell's theorem without inequalities are pertinent (see the end of Section \ref{sec:GHZ}).  Interestingly, in the present description the joint measurements considered there do not require kinks at all.  As $\theta_j=0$ for a measurement in the $x$ direction and $\theta_j=\pi/2 $ for the $y$ direction, in the case of a measurement of the $x$ components of all three spins, ($\theta_1=\theta_2=\theta_3=0$), Eq.~(\ref{eq:varphi_j+}) leads to $\sum_{j=1..3} \varphi_{j+} = 0$  only when the measurements outcomes satisfy $A_1A_2A_3=1$.  Similarly, in the three cases of two measured $y$ components and one $x$ component, $\sum_{j=1..3} \varphi_{j+} = 0$ only when $A_1A_2A_3=-1$. Outcomes of these sets of three measurements which would give the opposite value for the product correspond to $\sum_{j=1..3} \varphi_{j+} = \pi$. Therefore, given zero kinks (i.e., $\varphi_{j+}=\varphi_{j-}$ for all $j=1,2,3$), only the predicted outcome for the product is compatible with the initial condition $\varphi_{1-}+\varphi_{2-}+\varphi_{3-} = 0$. 
In light of Eq.~(\ref{eq:P(I,O,U)}) this means that these values of the product of the outcomes will appear with certainty, as necessary for the proof of Bell's theorem.

This example shows how the results of QM can be reproduced without violating the relevant type of locality condition.  This becomes possible due to the violation of the standard arrow-of-time condition: the influences which the late-time inputs, the values of the different settings $\theta_j$, have on the hidden variables associated with earlier times, $\varphi_{j-}$.  Note that knowledge of these $\varphi_{j-}$, the directions of the spins as they exit the source, determines the results of the measurements, $A_j$, and in this sense one might say that these results are ``predetermined,'' but here it is not at all implied that these predicted measurement results should be independent of the settings.  Thus, it does not follow that the model must supply values for $s_{jx}$ and $s_{jy}$ simultaneously.

\section{Elaboration}

In this section we discuss in detail the FID model, including its technicalities and extensions.
In the first subsection,
we demonstrate explicitly how the above description, including the ``as-few-kinks-as-possible'' rule, fits the local-FID stochastic framework of Eq.~(\ref{eq:P(I,O,U)}). 
In the second subsection, we present a quantum circuit that produces the general $N$-particle GHZ state, and explicitly provide the FID model parameters and rules associated with it's quantum gates and measurements. 

\subsection{Time discretization and weights}

Consider the as-few-as-possible kinks rule invoked above, which requires the $\vec q_j(t)$ to be 
piece-wise constant in time. 
This rule might appear to have a global character, but can be obtained within the framework of FID local models defined in Eq.~(\ref{eq:P(I,O,U)}) by identifying appropriate local weights (an analogy with the formation of large domains of essentially constant spin in a statistical-physics model of a ferromagnetic system with short-range interactions may be drawn).  It suffices to consider a single spin with a single final measurement, and discretize time so that the overall configuration $\vec q(t)$ is represented by a series of angles $\varphi_l$, with $l=0,1,\dots,L$.
The initial condition specifies $\varphi_0$, and the final measurement on that spin at an angle $\theta$ specifies that $\varphi_L$ is equal to either $\theta$ or $\theta+\pi$.  

The local weights
$w_l(\Delta \varphi)$ depend only on the differences between the angles at successive times, so that the overall weight for a specific history is $\prod_{l=1}^L w_l(\varphi_{l}-\varphi_{l-1})$.
The original Schulman model uses Lorentzians for the $w_l$, eventually taking their width parameter to zero.  This produces a L\'evy flight, with the vast majority of steps in the random walk having a negligible extent, and rare significant steps (or kinks) corresponding to the ``heavy tail'' of the distribution.
Here we suggest a simpler variant, with the two types of steps clearly segregated into continuous and discrete components, $w_l(\Delta\varphi) = w_{\rm C}(\Delta\varphi) + w_{\rm D}(\Delta\varphi)$, and with all angular variables discretized so that only values of $\varphi_m = \frac{2\pi m}{M}$ with $m = 1,2,\dots,M$ are allowed.  The discrete component is chosen as $w_{\rm D}(\Delta\varphi) = \delta_{\Delta\varphi,0}$ and the continuous component is $w_{\rm C}(\Delta\varphi) = \varepsilon \cos^2(\Delta \varphi /2)$, with the functional dependence chosen to reproduce Malus's law and the small parameter $\varepsilon$ ensuring that kinks are rare.
The required limit obtains when $\varepsilon LM \to 0$ and $M\rightarrow\infty$.

In this setup 
there is at most a single configuration with a constant $\varphi(t)$, with no kink and all $\varphi_l$ equal to each other, and its weight is $[w_{\rm D}(0)]^L = 1$. 
It exists if the initial and final conditions match, in the sense that $\varphi_0 = \varphi_L$ is possible.  For configurations with a single kink, there are $L$ possibilities for its timing, and two possible values of $\varphi_L$, given an initial $\varphi_0$ and a final measurement angle $\theta$.  There are thus $2L$ configurations, each with weight  $\varepsilon\cos^2 (\Delta \varphi / 2)$ where $\Delta \varphi = \varphi_L - \varphi_0$ due to the fact that $\varphi(t)$ is constant before and after the kink. 
Beyond that, there are $\frac{1}{2}L(L+1)$ possibilities for the timings of two kinks, with the corresponding weights proportional to $\varepsilon^2$.  With two kinks there is also a free choice of an intermediate angle among $M$ possibilities, so there are $ML(L+1)$ configurations.  For three kinks there are $O(L^3M^2)$ configurations, each with an $O(\varepsilon^3)$ weight, etc.
%
Using these weights in Eq.~\eqref{eq:P(I,O,U)} shows in explicit detail how the condition $\varepsilon LM \ll 1$ results in the rule of as-few-as-possible kinks.  

\subsection{GHZ states with $N$ spins}

\begin{figure}
\includegraphics[trim= 100 10 100  10, clip, scale=0.32]{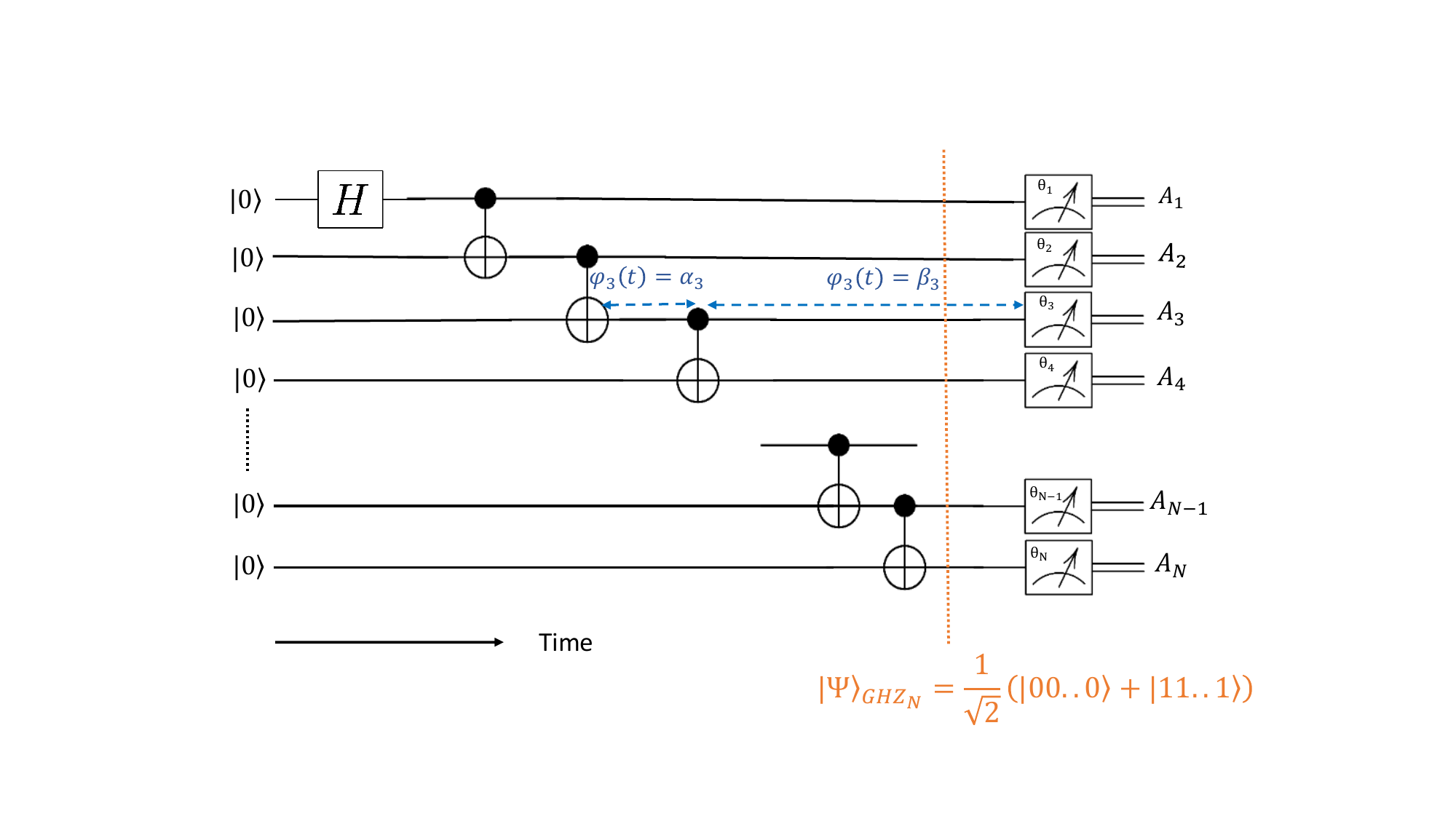}

\caption{Sketch of a quantum circuit of N qubits leading to the ${\rm GHZ}_{N}$ state (dotted line); also indicated are the subsequent measurements of each spin qubit in the $\theta_j$ direction in the $x-y$ plane, with outcome $A_j$.  $H$ denotes a Hadamard gate and each CNOT gate is depicted by a $\oplus$ sign on the affected bit connected to a dot on the control bit.  The quantum state is given in orange, and representative variables of the local-FID model, for the case wherein no kinks are needed, are indicated in blue.
\label{fig:A-GHZN-circuit}}
\end{figure}

We turn next to the GHZ $N$-spin analysis.  
The quantum circuit shown in Fig.~\ref{fig:A-GHZN-circuit} can be used to generate the initial ``all up $+$ all down'' quantum state for $N$ spins, or the ${\rm GHZ}_{N}$ state.  The circuit consists of $N$ qubits or spins, one Hadamard gate and $N-1$ CNOT gates.  Also included in the figure are the $N$ measurements performed on the $N$ spins at the final time, with the understanding that all measurements are made in the $x$-$y$ plane.
The predictions of QM for this setup are those of Eq.~(\ref{P_QM}).  We wish now to identify the hidden variables and the stochastic rules corresponding to the different elements of this circuit, in accordance with the description of the Schulman FID model above.

Note that each gate in the quantum circuit involves at most two spins, even when discussing a large-$N$ GHZ state.  In this sense, locality here, which is naturally defined by the gates, is more ``fine-grained'' than simply requiring 
that $\sum_j \varphi_{j0} = 0$.
As above, the description requires at most one kink for all of the $N$ spins involved.  It is easiest to begin with the case for which no kink at all is needed, because $\sum_{j=1}^N \theta_j$ is an integer multiple of $\pi$.  In this case, each of the angles $\varphi_j(t)$ will be constant along each of the time intervals of free propagation, away from the gates and the measurements.  One may introduce $\alpha_j$ as notation for the value of $\varphi_j(t)$ at times after the first gate for each spin, and $\beta_j$ for the values after the second gate and up to the measurement.
At time t=0 the value  of $\varphi_j$ is of no consequence, as all spins are in the $z$ direction (the model does not cover the times up to the first gate acting on each spin).
For $j = 1$ the Hadamard gate results in a spin in the $x$ direction, corresponding to $\alpha_1 = 0$. 
For the last spin, there is only a single gate, and one may take $\alpha_N = \beta_N$.  The final condition associated with the measurements implies $\beta_j = \theta_j + \pi (1-A_j)/2$.

It remains only to specify a rule for the CNOT gates. Denoting  $t_{Cj}$ the time of the CNOT gate between spin $j$ and $j+1$, with $t^-_{Cj}$, $t^+_{Cj}$  referring to times just before and after the gate, the suggested rule is 
\begin{equation} \label{eq:CNOT_rule}
\varphi_{j+1}\left(t^+_{Cj}\right)  + \varphi_j\left(t^+_{Cj}\right) = \varphi_{j}\left(t^-_{Cj}\right).
\end{equation}
In other words, the sum of the angles after each CNOT gate is equal to the single well-defined angle before it. This corresponds in the FID model to a local weight of the variables near the CNOT gate, $w_{\rm C}(\varphi^{\rm C+}_j,\varphi^{\rm C+}_{j+1},\varphi^{\rm C-}_j) \propto \delta(\varphi^{\rm C+}_{j+1} + \varphi^{\rm C+}_j - \varphi^{\rm C-}_j)$.  Translating to the discrete angles used here, this implies 
$\alpha_{j+1}  + \beta_j = \alpha_{j}$.

The consequences of this rule are $\alpha_2 = -\beta_1$ for the first CNOT gate,  $\alpha_3 = -\beta_1-\beta_2$ for the second, and so on up to  $\alpha_N = -\sum_{j=1}^{N-1} \beta_j$.  
Since $\alpha_N=\beta_N$, we find that the rules invoked for the CNOT gates and the Hadamard gate impose the overall condition $\sum_{j=1}^N \beta_j = 0$. 
These rules suffice to reproduce the GHZ correlations for measurements with 
maximal correlations.  Note 
that there are $2^{N-1}$ possible outcomes with 
these
correlations, all corresponding to different configurations in our model with equal weights and probabilities.

Finally, consider the case in which $\sum_{j=1}^N \theta_j$ is not an integer multiple of $\pi$.  From the above discussion, in this case at least one kink is needed, and it can occur anywhere between the gates and the measurements. Following the FID rules above, one finds that the phase shift of the kink $\Delta\varphi$ must have a weight of $w_k \propto \cos^2\frac12\Delta\varphi$. The sum over all well-defined angles of all the spins up to the time of the kink $t_k$ is $0$, and after $t_k$ it is $\Delta \varphi$. The phase shift is therefore related to the measurements outcomes by
\begin{equation}    \Delta\varphi=\sum_{j=1}^N{\theta_j+\frac{\pi}{2}\left(1-A_j\right)} \; .
\end{equation}
Summing up all the histories which are selected by the above rules, with their defined local positive weights and with the normalization according to Eq.~(\ref{eq:P(I,O,U)}), indeed produces precisely the probabilities of 
Eq.~(\ref{P_QM}).


%

\section{Discussion and outlook}

It was shown above that the Schulman model can be extended to cover GHZ correlations, for three spins or more, with only minor modifications to its rules.  The FID approach thus provides ``reasonable'' descriptions not only for violations of the Leggett-Garg inequality \cite{leggett1985} involving repeated measurements of a single spin, and violations of Bell inequalities involving two spins \cite{wharton2020}, but also for the quantum behavior of the three-spin systems used in proving quantum non-locality without inequalities \cite{mermin1990}.

The word ``reasonable'' here signifies that these are stochastic models employing the standard probability rules, and that they are local and causal in appropriate senses.  The locality condition we have used here, that of Gibbs fields with short-ranged weights, Eq.~\eqref{eq:P(I,O,U)}, differs from Bell's locality condition, but the models discussed here fulfill Bell's condition as well.  More generally, any local-Gibbs-field model will obviously fulfill the Continuous Action condition of Ref.~\cite{wharton2020}, which generalizes the ``screening condition,'' which Bell used to define locality, to situations with FID.
Regarding causality, although FID models are obviously non-standard, the presented models are in line with the following physical expectation: they are time-symmetric at the most basic level, with the time-symmetry broken by the initial and final conditions in a manner leading to signal causality.  True, these models may appear perplexing due to the violation of the standard causal arrow of time, but viewing this as a disadvantage may well be just the type of prejudice which should be avoided.

There are many features of FID models which deserve discussion --- the interested reader is referred, again, to the review~\cite{wharton2020}.  One point bears repetition: in this type of model, the actual state of the system at time $t$, which consists of the values of the local variables $\varphi_j(t)$ at that time, is only indirectly related to its quantum state. %
In contrast, the quantum state $\Psi$ at a time $t$ is in this view a mathematical tool representing the available information 
up to that time.  No wonder that $\Psi$ has peculiar features:
\begin{enumerate}[topsep=3pt,itemsep=2pt,label=(\alph*)]
    \item It is abstract (a ray in Hilbert space);
    \item For large multi-component systems, it's complexity grows  exponentially 
with the number of components
(like the phase-space distributions of the Liouville equation);
    \item It preserves information (unitary evolution) between times $t$ and $t'$ if there is no change in the available information over the time interval $(t,t')$;
    \item It changes abruptly (collapse) when a measurement is made and its result is irreversibly registered and becomes available;
    \item It requires the hypersurface simultaneous with $t$ to be well-defined, clashing with relativity (at least in concept).
\end{enumerate}

Beyond the situations addressed by the above extended Schulman model, there are many scenarios for which an FID, continuous-action model of this type is not yet available, not even in a toy-model version.  Examples include not only more-general quantum circuits but also simple systems such as the GHZ state with three spin measurements not in the  $x$-$y$ plane, and the triangle setup of \textcite{renou2019}.  Furthermore, the rules of the above model are rather ad hoc, especially those pertaining to measurements.  There are thus plenty of challenges ahead.  Perhaps a switch from a ``particles description'' to a ``fields'' point of view is necessary for significant progress to be made \cite{hobson2013}.  Time will tell.


\bibliography{refs}

\begin{thebibliography}{22}%
\makeatletter
\providecommand \@ifxundefined [1]{%
 \@ifx{#1\undefined}
}%
\providecommand \@ifnum [1]{%
 \ifnum #1\expandafter \@firstoftwo
 \else \expandafter \@secondoftwo
 \fi
}%
\providecommand \@ifx [1]{%
 \ifx #1\expandafter \@firstoftwo
 \else \expandafter \@secondoftwo
 \fi
}%
\providecommand \natexlab [1]{#1}%
\providecommand \enquote  [1]{``#1''}%
\providecommand \bibnamefont  [1]{#1}%
\providecommand \bibfnamefont [1]{#1}%
\providecommand \citenamefont [1]{#1}%
\providecommand \href@noop [0]{\@secondoftwo}%
\providecommand \href [0]{\begingroup \@sanitize@url \@href}%
\providecommand \@href[1]{\@@startlink{#1}\@@href}%
\providecommand \@@href[1]{\endgroup#1\@@endlink}%
\providecommand \@sanitize@url [0]{\catcode `\\12\catcode `\$12\catcode
  `\&12\catcode `\#12\catcode `\^12\catcode `\_12\catcode `\%12\relax}%
\providecommand \@@startlink[1]{}%
\providecommand \@@endlink[0]{}%
\providecommand \url  [0]{\begingroup\@sanitize@url \@url }%
\providecommand \@url [1]{\endgroup\@href {#1}{\urlprefix }}%
\providecommand \urlprefix  [0]{URL }%
\providecommand \Eprint [0]{\href }%
\providecommand \doibase [0]{http://dx.doi.org/}%
\providecommand \selectlanguage [0]{\@gobble}%
\providecommand \bibinfo  [0]{\@secondoftwo}%
\providecommand \bibfield  [0]{\@secondoftwo}%
\providecommand \translation [1]{[#1]}%
\providecommand \BibitemOpen [0]{}%
\providecommand \bibitemStop [0]{}%
\providecommand \bibitemNoStop [0]{.\EOS\space}%
\providecommand \EOS [0]{\spacefactor3000\relax}%
\providecommand \BibitemShut  [1]{\csname bibitem#1\endcsname}%
\let\auto@bib@innerbib\@empty
\bibitem [{\citenamefont {Einstein}\ \emph {et~al.}(1935)\citenamefont
  {Einstein}, \citenamefont {Podolsky},\ and\ \citenamefont {Rosen}}]{EPR1935}%
  \BibitemOpen
  \bibfield  {author} {\bibinfo {author} {\bibfnamefont {A.}~\bibnamefont
  {Einstein}}, \bibinfo {author} {\bibfnamefont {B.}~\bibnamefont {Podolsky}},
  \ and\ \bibinfo {author} {\bibfnamefont {N.}~\bibnamefont {Rosen}},\
  }\bibfield  {title} {\enquote {\bibinfo {title} {Can quantum-mechanical
  description of physical reality be considered complete?}}\ }\href {\doibase
  10.1103/PhysRev.47.777} {\bibfield  {journal} {\bibinfo  {journal} {Phys.
  Rev.}\ }\textbf {\bibinfo {volume} {47}},\ \bibinfo {pages} {777--780}
  (\bibinfo {year} {1935})}\BibitemShut {NoStop}%
\bibitem [{\citenamefont {Bell}(2004)}]{bell2004}%
  \BibitemOpen
  \bibfield  {author} {\bibinfo {author} {\bibfnamefont {J.~S.}\ \bibnamefont
  {Bell}},\ }\href {\doibase 10.1017/CBO9780511815676} {\emph {\bibinfo {title}
  {Speakable and Unspeakable in Quantum Mechanics: Collected Papers on Quantum
  Philosophy}}},\ \bibinfo {edition} {2nd}\ ed.\ (\bibinfo  {publisher}
  {Cambridge University Press},\ \bibinfo {year} {2004})\BibitemShut {NoStop}%
\bibitem [{\citenamefont {'t~Hooft}(2016)}]{thooft2016}%
  \BibitemOpen
  \bibfield  {author} {\bibinfo {author} {\bibfnamefont {Gerard}\ \bibnamefont
  {'t~Hooft}},\ }\href {\doibase 10.1007/978-3-319-41285-6} {\emph {\bibinfo
  {title} {The cellular automaton interpretation of quantum mechanics}}}\
  (\bibinfo  {publisher} {Springer},\ \bibinfo {year} {2016})\BibitemShut
  {NoStop}%
\bibitem [{\citenamefont {Waegell}\ and\ \citenamefont
  {McQueen}(2020)}]{waegell2020}%
  \BibitemOpen
  \bibfield  {author} {\bibinfo {author} {\bibfnamefont {Mordecai}\
  \bibnamefont {Waegell}}\ and\ \bibinfo {author} {\bibfnamefont {Kelvin~J.}\
  \bibnamefont {McQueen}},\ }\bibfield  {title} {\enquote {\bibinfo {title}
  {Reformulating bell's theorem: The search for a truly local quantum
  theory},}\ }\href {\doibase https://doi.org/10.1016/j.shpsb.2020.02.006}
  {\bibfield  {journal} {\bibinfo  {journal} {Studies in History and Philosophy
  of Science Part B: Studies in History and Philosophy of Modern Physics}\
  }\textbf {\bibinfo {volume} {70}},\ \bibinfo {pages} {39--50} (\bibinfo
  {year} {2020})}\BibitemShut {NoStop}%
\bibitem [{\citenamefont {Price}(1997)}]{price1997}%
  \BibitemOpen
  \bibfield  {author} {\bibinfo {author} {\bibfnamefont {Huw}\ \bibnamefont
  {Price}},\ }\href
  {https://global.oup.com/academic/product/times-arrow-and-archimedes-point-9780195117981?cc=gb&lang=en&#}
  {\emph {\bibinfo {title} {Time's arrow \& Archimedes' point: new directions
  for the physics of time}}}\ (\bibinfo  {publisher} {Oxford University Press,
  USA},\ \bibinfo {year} {1997})\BibitemShut {NoStop}%
\bibitem [{\citenamefont {Wharton}(2015)}]{wharton2015a}%
  \BibitemOpen
  \bibfield  {author} {\bibinfo {author} {\bibfnamefont {Ken}\ \bibnamefont
  {Wharton}},\ }\bibfield  {title} {\enquote {\bibinfo {title} {The universe is
  not a computer},}\ }in\ \href
  {https://link.springer.com/book/10.1007/978-3-319-13045-3} {\emph {\bibinfo
  {booktitle} {Questioning the foundations of physics}}},\ \bibinfo {editor}
  {edited by\ \bibinfo {editor} {\bibfnamefont {Anthony}\ \bibnamefont
  {Aguirre}}, \bibinfo {editor} {\bibfnamefont {Brendan}\ \bibnamefont
  {Foster}}, \ and\ \bibinfo {editor} {\bibfnamefont {Zeeya}\ \bibnamefont
  {Merali}}}\ (\bibinfo  {publisher} {Springer},\ \bibinfo {year} {2015})\ pp.\
  \bibinfo {pages} {177--189}\BibitemShut {NoStop}%
\bibitem [{\citenamefont {Wharton}\ and\ \citenamefont
  {Argaman}(2020)}]{wharton2020}%
  \BibitemOpen
  \bibfield  {author} {\bibinfo {author} {\bibfnamefont {K.~B.}\ \bibnamefont
  {Wharton}}\ and\ \bibinfo {author} {\bibfnamefont {N.}~\bibnamefont
  {Argaman}},\ }\bibfield  {title} {\enquote {\bibinfo {title} {Colloquium:
  Bell's theorem and locally mediated reformulations of quantum mechanics},}\
  }\href {\doibase 10.1103/RevModPhys.92.021002} {\bibfield  {journal}
  {\bibinfo  {journal} {Reviews of Modern Physics}\ }\textbf {\bibinfo {volume}
  {92}},\ \bibinfo {pages} {021002} (\bibinfo {year} {2020})}\BibitemShut
  {NoStop}%
\bibitem [{\citenamefont {Bell}(1990)}]{bell1990}%
  \BibitemOpen
  \bibfield  {author} {\bibinfo {author} {\bibfnamefont {JS}~\bibnamefont
  {Bell}},\ }\bibfield  {title} {\enquote {\bibinfo {title} {La nouvelle
  cuisine},}\ }in\ \href@noop {} {\emph {\bibinfo {booktitle} {Between Science
  and Technology}}},\ \bibinfo {editor} {edited by\ \bibinfo {editor}
  {\bibfnamefont {A}~\bibnamefont {Sarlemijn}}\ and\ \bibinfo {editor}
  {\bibfnamefont {P}~\bibnamefont {Kroes}}}\ (\bibinfo  {publisher}
  {Elsevier},\ \bibinfo {year} {1990})\ pp.\ \bibinfo {pages}
  {97--115}\BibitemShut {NoStop}%
\bibitem [{\citenamefont {Hill}(2022)}]{hill2022}%
  \BibitemOpen
  \bibfield  {author} {\bibinfo {author} {\bibfnamefont {Heather~M.}\
  \bibnamefont {Hill}},\ }\bibfield  {title} {\enquote {\bibinfo {title}
  {{Physics Nobel honors foundational quantum entanglement experiments}},}\
  }\href {\doibase 10.1063/PT.3.5133} {\bibfield  {journal} {\bibinfo
  {journal} {Physics Today}\ }\textbf {\bibinfo {volume} {75}},\ \bibinfo
  {pages} {14--17} (\bibinfo {year} {2022})}\BibitemShut {NoStop}%
\bibitem [{\citenamefont {Greenberger}\ \emph {et~al.}(1989)\citenamefont
  {Greenberger}, \citenamefont {Horne},\ and\ \citenamefont {Zeilinger}}]{GHZ}%
  \BibitemOpen
  \bibfield  {author} {\bibinfo {author} {\bibfnamefont {Daniel~M.}\
  \bibnamefont {Greenberger}}, \bibinfo {author} {\bibfnamefont {Michael~A.}\
  \bibnamefont {Horne}}, \ and\ \bibinfo {author} {\bibfnamefont {Anton}\
  \bibnamefont {Zeilinger}},\ }\enquote {\bibinfo {title} {Going beyond
  {B}ell's theorem},}\ in\ \href {\doibase 10.1007/978-94-017-0849-4_10} {\emph
  {\bibinfo {booktitle} {Bell's Theorem, Quantum Theory and Conceptions of the
  Universe}}},\ \bibinfo {editor} {edited by\ \bibinfo {editor} {\bibfnamefont
  {Menas}\ \bibnamefont {Kafatos}}}\ (\bibinfo  {publisher} {Springer
  Netherlands},\ \bibinfo {address} {Dordrecht},\ \bibinfo {year} {1989})\ pp.\
  \bibinfo {pages} {69--72}\BibitemShut {NoStop}%
\bibitem [{\citenamefont {Greenberger}\ \emph {et~al.}(1990)\citenamefont
  {Greenberger}, \citenamefont {Horne}, \citenamefont {Shimony},\ and\
  \citenamefont {Zeilinger}}]{GHSZ1990}%
  \BibitemOpen
  \bibfield  {author} {\bibinfo {author} {\bibfnamefont {Daniel~M.}\
  \bibnamefont {Greenberger}}, \bibinfo {author} {\bibfnamefont {Michael~A.}\
  \bibnamefont {Horne}}, \bibinfo {author} {\bibfnamefont {Abner}\ \bibnamefont
  {Shimony}}, \ and\ \bibinfo {author} {\bibfnamefont {Anton}\ \bibnamefont
  {Zeilinger}},\ }\bibfield  {title} {\enquote {\bibinfo {title} {Bell’s
  theorem without inequalities},}\ }\href {\doibase 10.1119/1.16243} {\bibfield
   {journal} {\bibinfo  {journal} {American Journal of Physics}\ }\textbf
  {\bibinfo {volume} {58}},\ \bibinfo {pages} {1131--1143} (\bibinfo {year}
  {1990})},\ \Eprint {http://arxiv.org/abs/https://doi.org/10.1119/1.16243}
  {https://doi.org/10.1119/1.16243} \BibitemShut {NoStop}%
\bibitem [{\citenamefont {Mermin}(1990)}]{mermin1990}%
  \BibitemOpen
  \bibfield  {author} {\bibinfo {author} {\bibfnamefont {N.~David}\
  \bibnamefont {Mermin}},\ }\bibfield  {title} {\enquote {\bibinfo {title}
  {Quantum mysteries revisited},}\ }\href {\doibase 10.1119/1.16503} {\bibfield
   {journal} {\bibinfo  {journal} {American Journal of Physics}\ }\textbf
  {\bibinfo {volume} {58}},\ \bibinfo {pages} {731--734} (\bibinfo {year}
  {1990})},\ \Eprint {http://arxiv.org/abs/https://doi.org/10.1119/1.16503}
  {https://doi.org/10.1119/1.16503} \BibitemShut {NoStop}%
\bibitem [{\citenamefont {Schulman}(1997)}]{schulman1997}%
  \BibitemOpen
  \bibfield  {author} {\bibinfo {author} {\bibfnamefont {Lawrence~S}\
  \bibnamefont {Schulman}},\ }\href
  {https://www.cambridge.org/us/universitypress/subjects/physics/history-philosophy-and-foundations-physics/times-arrows-and-quantum-measurement?format=HB&isbn=9780521561228}
  {\emph {\bibinfo {title} {Time's arrows and quantum measurement}}}\ (\bibinfo
   {publisher} {Cambridge University Press},\ \bibinfo {year}
  {1997})\BibitemShut {NoStop}%
\bibitem [{\citenamefont {Wharton}(2014)}]{wharton2014}%
  \BibitemOpen
  \bibfield  {author} {\bibinfo {author} {\bibfnamefont {Ken}\ \bibnamefont
  {Wharton}},\ }\bibfield  {title} {\enquote {\bibinfo {title} {Quantum states
  as ordinary information},}\ }\href {https://www.mdpi.com/2078-2489/5/1/190}
  {\bibfield  {journal} {\bibinfo  {journal} {Information}\ }\textbf {\bibinfo
  {volume} {5}},\ \bibinfo {pages} {190--208} (\bibinfo {year}
  {2014})}\BibitemShut {NoStop}%
\bibitem [{\citenamefont {Wharton}(2016)}]{wharton2016}%
  \BibitemOpen
  \bibfield  {author} {\bibinfo {author} {\bibfnamefont {Ken}\ \bibnamefont
  {Wharton}},\ }\bibfield  {title} {\enquote {\bibinfo {title} {Towards a
  realistic parsing of the {F}eynman {P}ath {I}ntegral},}\ }\href
  {http://quanta.ws/ojs/index.php/quanta/article/view/41} {\bibfield  {journal}
  {\bibinfo  {journal} {Quanta}\ }\textbf {\bibinfo {volume} {5}},\ \bibinfo
  {pages} {1--11} (\bibinfo {year} {2016})}\BibitemShut {NoStop}%
\bibitem [{\citenamefont {Leggett}\ and\ \citenamefont
  {Garg}(1985)}]{leggett1985}%
  \BibitemOpen
  \bibfield  {author} {\bibinfo {author} {\bibfnamefont {A.~J.}\ \bibnamefont
  {Leggett}}\ and\ \bibinfo {author} {\bibfnamefont {Anupam}\ \bibnamefont
  {Garg}},\ }\bibfield  {title} {\enquote {\bibinfo {title} {Quantum mechanics
  versus macroscopic realism: Is the flux there when nobody looks?}}\ }\href
  {\doibase 10.1103/PhysRevLett.54.857} {\bibfield  {journal} {\bibinfo
  {journal} {Phys. Rev. Lett.}\ }\textbf {\bibinfo {volume} {54}},\ \bibinfo
  {pages} {857--860} (\bibinfo {year} {1985})}\BibitemShut {NoStop}%
\bibitem [{\citenamefont {Maudlin}(1994)}]{maudlin1994}%
  \BibitemOpen
  \bibfield  {author} {\bibinfo {author} {\bibfnamefont {Tim}\ \bibnamefont
  {Maudlin}},\ }\href@noop {} {\emph {\bibinfo {title} {Quantum Non‐Locality
  and Relativity}}},\ \bibinfo {edition} {1st}\ ed.\ (\bibinfo  {publisher}
  {Blackwell Publishing},\ \bibinfo {year} {1994})\BibitemShut {NoStop}%
\bibitem [{\citenamefont {Wood}\ and\ \citenamefont
  {Spekkens}(2015)}]{wood2015}%
  \BibitemOpen
  \bibfield  {author} {\bibinfo {author} {\bibfnamefont {Christopher~J}\
  \bibnamefont {Wood}}\ and\ \bibinfo {author} {\bibfnamefont {Robert~W}\
  \bibnamefont {Spekkens}},\ }\bibfield  {title} {\enquote {\bibinfo {title}
  {The lesson of causal discovery algorithms for quantum correlations: Causal
  explanations of {B}ell-inequality violations require fine-tuning},}\ }\href
  {https://iopscience.iop.org/article/10.1088/1367-2630/17/3/033002} {\bibfield
   {journal} {\bibinfo  {journal} {New Journal of Physics}\ }\textbf {\bibinfo
  {volume} {17}},\ \bibinfo {pages} {033002} (\bibinfo {year}
  {2015})}\BibitemShut {NoStop}%
\bibitem [{\citenamefont {Berkovitz}(2008)}]{berkovitz2008}%
  \BibitemOpen
  \bibfield  {author} {\bibinfo {author} {\bibfnamefont {Joseph}\ \bibnamefont
  {Berkovitz}},\ }\bibfield  {title} {\enquote {\bibinfo {title} {On
  predictions in retro-causal interpretations of quantum mechanics},}\ }\href
  {\doibase 10.1016/j.shpsb.2008.08.002} {\bibfield  {journal} {\bibinfo
  {journal} {Studies in History and Philosophy of Science Part B: Studies in
  History and Philosophy of Modern Physics}\ }\textbf {\bibinfo {volume}
  {39}},\ \bibinfo {pages} {709 -- 735} (\bibinfo {year} {2008})}\BibitemShut
  {NoStop}%
\bibitem [{\citenamefont {Almada}\ \emph {et~al.}(2016)\citenamefont {Almada},
  \citenamefont {Ch'ng}, \citenamefont {Kintner}, \citenamefont {Morrison},\
  and\ \citenamefont {Wharton}}]{almada2016}%
  \BibitemOpen
  \bibfield  {author} {\bibinfo {author} {\bibfnamefont {D}~\bibnamefont
  {Almada}}, \bibinfo {author} {\bibfnamefont {K}~\bibnamefont {Ch'ng}},
  \bibinfo {author} {\bibfnamefont {S}~\bibnamefont {Kintner}}, \bibinfo
  {author} {\bibfnamefont {B}~\bibnamefont {Morrison}}, \ and\ \bibinfo
  {author} {\bibfnamefont {KB}~\bibnamefont {Wharton}},\ }\bibfield  {title}
  {\enquote {\bibinfo {title} {Are retrocausal accounts of entanglement
  unnaturally fine-tuned?}}\ }\href {https://ijqf.org/archives/3218} {\bibfield
   {journal} {\bibinfo  {journal} {International Journal of Quantum
  Foundations}\ }\textbf {\bibinfo {volume} {2}},\ \bibinfo {pages} {1--14}
  (\bibinfo {year} {2016})}\BibitemShut {NoStop}%
\bibitem [{\citenamefont {Renou}\ \emph {et~al.}(2019)\citenamefont {Renou},
  \citenamefont {B\"aumer}, \citenamefont {Boreiri}, \citenamefont {Brunner},
  \citenamefont {Gisin},\ and\ \citenamefont {Beigi}}]{renou2019}%
  \BibitemOpen
  \bibfield  {author} {\bibinfo {author} {\bibfnamefont {Marc-Olivier}\
  \bibnamefont {Renou}}, \bibinfo {author} {\bibfnamefont {Elisa}\ \bibnamefont
  {B\"aumer}}, \bibinfo {author} {\bibfnamefont {Sadra}\ \bibnamefont
  {Boreiri}}, \bibinfo {author} {\bibfnamefont {Nicolas}\ \bibnamefont
  {Brunner}}, \bibinfo {author} {\bibfnamefont {Nicolas}\ \bibnamefont
  {Gisin}}, \ and\ \bibinfo {author} {\bibfnamefont {Salman}\ \bibnamefont
  {Beigi}},\ }\bibfield  {title} {\enquote {\bibinfo {title} {Genuine quantum
  nonlocality in the triangle network},}\ }\href {\doibase
  10.1103/PhysRevLett.123.140401} {\bibfield  {journal} {\bibinfo  {journal}
  {Phys. Rev. Lett.}\ }\textbf {\bibinfo {volume} {123}},\ \bibinfo {pages}
  {140401} (\bibinfo {year} {2019})}\BibitemShut {NoStop}%
\bibitem [{\citenamefont {Hobson}(2013)}]{hobson2013}%
  \BibitemOpen
  \bibfield  {author} {\bibinfo {author} {\bibfnamefont {Art}\ \bibnamefont
  {Hobson}},\ }\bibfield  {title} {\enquote {\bibinfo {title} {There are no
  particles, there are only fields},}\ }\href {\doibase 10.1119/1.4789885}
  {\bibfield  {journal} {\bibinfo  {journal} {American Journal of Physics}\
  }\textbf {\bibinfo {volume} {81}},\ \bibinfo {pages} {211--223} (\bibinfo
  {year} {2013})},\ \Eprint
  {http://arxiv.org/abs/https://doi.org/10.1119/1.4789885}
  {https://doi.org/10.1119/1.4789885} \BibitemShut {NoStop}%
\end{thebibliography}%
\end{document}